\documentclass[10pt]{artikel3}
\usepackage{amsfonts,amsmath,amssymb}
\usepackage{float,braket}
\usepackage{times}
\usepackage{subfigure}
\usepackage{graphicx,wrapfig,color}

\newcommand{\p}{\partial}

\usepackage[small,bf]{caption}
\newcommand{\sect}[1]{ \section{#1} \setcounter{equation}{0} }

\definecolor{Rood}{rgb}{1, 0, 0} 

\begin{document}
\title{{\bf On the computation of the spectral density of two-point functions: complex masses, cut rules and beyond}}
\author{David Dudal$^{a}$\thanks{david.dudal@ugent.be} ~\& Marcelo S.~Guimaraes$^b$\thanks{msguimaraes@uerj.br}\\\\
\small \textnormal{$^{a}$ Ghent University, Department of Physics and Astronomy, Krijgslaan 281-S9, 9000 Gent, Belgium}
\\
\small \textnormal{$^{b}$ Departamento de F\'{\i }sica Te\'{o}rica, Instituto de F\'{\i }sica, UERJ - Universidade do Estado do Rio de Janeiro,}
\\ \small \textnormal{\phantom{$^{b}$} Rua S\~{a}o Francisco Xavier 524, 20550-013 Maracan\~{a}, Rio de Janeiro, Brasil}\normalsize}

\date{}
\maketitle
\begin{abstract}
We present a steepest descent calculation of the K\"all\'{e}n-Lehmann spectral density of two-point functions involving complex conjugate masses in Euclidean space. This problem occurs in studies of (gauge) theories with Gribov-like propagators. As the presence of complex masses and the use of Euclidean space brings the theory outside of the strict validity of the Cutkosky cut rules, we discuss an alternative method based on the Widder inversion operator of the Stieltjes transformation. It turns out that the results coincide with those obtained by naively applying the cut rules. We also point out the potential usefulness of the Stieltjes (inversion) formalism when non-standard propagators are used, in which case cut rules are not available at all.
\end{abstract}

\setcounter{page}{1}

\sect{Introduction}
In the setting of local quantum field theory, all information on the spectrum resides in the Green functions of the theory, as they build up the scattering $\mathcal{S}$-matrix. A special class of Green functions are the two-point functions. We should not only consider the elementary propagators of the theory, but also the two-point functions of local composite operators, which correspond to propagating bound states, are relevant.

Scattering amplitudes are in general analytic functions of the Lorentz invariants constructed from the external momenta characterizing a given process. In the present work we shall only be interested in two-point functions which can be considered as functions of the external momentum $k^2=s$, analytically continued to the complex $s$-plane with a cut on the real axis starting at $s=\tau_0\geq 0$, which is identified as the threshold for multi-particle production. Such a two-point correlation function can be cast in the form of a dispersion relation, that is, an integral representation written in terms of the function's discontinuities in the complex plane. For the two-point function of an arbitrary operator ${\cal O}$ we can write the general expression
\begin{align}
\braket{{\cal O}(k) {\cal O}(-k)} = {\cal F} (s) = \frac{1}{\pi}\int^{\infty}_{0} d\tau \; \frac{\text{Im } {\cal F} (\tau)}{\tau - s}\;. \label{cutk-drel}
\end{align}
The imaginary part in the argument of the integral stands for the discontinuity of the function ${\cal F} (s)$ across the cut. This ought to be a positive definite quantity proportional to the total cross-section, as demanded by the optical theorem. The representation \eqref{cutk-drel} is known as the K\"all\'{e}n-Lehmann spectral representation. For a concise treatment, let us refer to e.g.~ \cite{smatrix1,smatrix2,Itzykson:1980rh,Peskin:1995ev}.

This K\"all\'{e}n-Lehmann spectral representation can be constructed, for the case of real masses in Minkowski space, through the use of the Cutkosky cut rules \cite{smatrix2,Cutkosky:1960sp}, which state that, in order to calculate the discontinuity associated with a Feynman diagram, we just need to cut propagator lines in the diagram in all possible ways, thereby replacing the corresponding propagators by $\delta$-functions localizing to the physical phase space, and then to sum over all these contributions. Physically, we place the particles running in these lines on-shell.

Specifying for a moment to the strong interaction, it is well-known that QCD exhibits confinement, i.e.~the color-charged elementary degrees of freedom (quarks and gluons) do not propagate in the physical world. The observables are the color-neutral bound states, corresponding to the gauge invariant operators constructed from the gluon and quarks fields, being the mesons, baryons, glueballs and quark-gluon hybrids. A big contemporary challenge is to extract information on that nonperturbative spectrum using theoretical, phenomenological and/or numerical lattice tools. In the so-called sum rules approach to the QCD spectrum, spectral representations play a pivotal role, we refer to the comprehensive tome \cite{Narison:2002pw}, to \cite{Shifman:1978bx,Shifman:1978by} for the seminal works, or to \cite{Peskin:1995ev} for a pedestrian's introduction.

In many approaches to QCD, including the sum rules, an important role is played by the elementary propagators, in particular the gluon propagator, which is usually studied in the Landau gauge, $\p_\mu A_\mu=0$. The elementary propagators are the fundamental building blocks of more complicated Green functions, including the ones corresponding to bound states \cite{Alkofer:2000wg,Roberts:1994dr,Roberts:2007ji}. For some recent advancements on the gluon propagator, including numerical lattice results, let us refer to the non-extensive list of \cite{Cucchieri:2007md,Cucchieri:2007mdb,Cucchieri:2007mdc,Fischer:2008uz,Fischer:2008uzb,Fischer:2008uzc,Dudal:2008sp,Dudal:2010tf,Gracey:2010cg,Boucaud:2010gr,Tissier:2010ts,Furui:2004cx,Maas:2009se,Kondo:2009ug,Iritani:2009mp,Bogolubsky:2009dc,Oliveira:2010xc}.

It is apparent that the Landau gauge gluon propagator exhibits a violation of positivity \cite{Osterwalder:1974tc,Osterwalder:1974tcb}, a fact which also received direct lattice confirmation \cite{Silva:2006bs,Bowman:2007du}. This violation of positivity means nothing more than that there cannot exist a K\"all\'{e}n-Lehmann representation \cite{Roberts:2007ji}, hence the Landau gauge gluon is unphysical, and it can be interpreted as a reflection of confinement. In \cite{Dudal:2008sp,Dudal:2010tf}, it was discussed how
\begin{equation}\label{prop}
D(k^2)=\frac{k^2+M^2}{k^4+(M^2+m^2)k^2+\lambda^4}
\end{equation}
can qualitatively and quantitatively describe the lattice predictions for the (infrared) Landau gauge gluon propagator, and this up to $k\sim 1.5~\text{GeV}$. The expression \eqref{prop} can be seen as a dynamically improved version of the so-called Gribov propagator,
\begin{equation}\label{prop2}
D(k^2)=\frac{k^2}{k^4+\lambda^4}\,,
\end{equation}
which attracted a lot of attention as a prototype of a confining gluon propagator, and which is related to the dealing with the issue of gauge (Gribov) copies in the Landau gauge \cite{Gribov:1977wm,Zwanziger:1989mf,Zwanziger:1989mfb}. Given that \eqref{prop} works out well, it might subsequently be used in explicit computations, for example when one would like to learn something on the glueball or meson spectrum.

We notice that the propagator \eqref{prop} can be decomposed into 2 Yukawa propagators, with either 2 real, or 2 complex conjugate masses, depending on the relative size of the scales present. The case of 2 complex conjugate masses seems to be realized \cite{Dudal:2010tf}, and a (tree level) positivity violation is then guaranteed. The unphysical degrees of freedom corresponding to those complex conjugate masses were called $i$-particles \cite{Baulieu:2009ha}. Also the so-called Stingl propagator \cite{Stingl:1985hx,Stingl:1994nk},
\begin{equation}\label{prop3}
D(k^2)=\frac{k^2}{k^4+m^2k^2+\lambda^4}\,,
\end{equation}
can exhibit a pair of complex conjugate poles, as a special case of \eqref{prop}.

If we are interested in a spectral function inspired approach to e.g.~bound states in theories with propagators of the type \eqref{prop}, we of course would need to know how to compute with such a propagator. Research in this direction was initiated with the work on a toy model \cite{Baulieu:2009ha}, where the necessary spectral densities were computed by clever manipulations, which unfortunately do not have general applicability, or, more realistically, for gauge theories in \cite{Capri:2010pg,Dudal:2010cd} by using the Cutkosky cut rules and assuming that the results can be continued to Euclidean space and the case of complex conjugate masses.

The main purpose of this paper is to show that the cut rules results, which we summarized in Section 2, are actually correct. We shall therefore bring to attention in Section 3 the Stieltjes integral transformation, as well as its inverse, as studied by Widder \cite{stieltjes1}, which allows to compute directly in Euclidean space the spectral density, even in cases when there is no cut rules prescription. We shall work out an introductory example in Section 4. In Section 5, we comment on the use of subtracted spectral representations to avoid divergences. We then present in Section 6 in full detail the most interesting case of 2 $i$-particle propagators with complex conjugate masses, wherefore a completely analytical treatment can also be given. In our concluding remarks, Section 7, we point out that the Widder result to compute the spectral function could perhaps also be used in a numerical context when more complicated propagators than e.g.~\eqref{prop} are used, which are completely inadequate for fully analytical computations.

\sect{Using the Cutkosky cut rules}
\subsection{Preliminaries}
We assume that $\mathcal{O}$ is a generic (composite) operator containing 2 fields. We are then interested in the study of the following $1$-loop expression
\begin{align}
\Braket{ {\cal O}(k) {\cal O}(-k)} \rightarrow {\cal F}(k,m_1,m_2) = \int \frac{d^dp}{(2\pi)^d}\; \frac{1}{(k-p)^2-m_1^2} \frac{1}{p^2-m_2^2} f(p,k-p)\,. \label{1-loop-mink}
\end{align}
where $f(p,k-p)$ is a function of the Lorentz invariants constructed from the momenta flowing in the loop. For simplicity, we may assume $\cal{O}$ to be a scalar function, thus a fortiori also $f(p,k-p)$. If $\cal{O}$ would carry Lorentz indices, say $\mathcal{O}=\mathcal{O}_{\mu_1\ldots\mu_n}$, the eventual correlation function $\Braket{\mathcal{O}_{\mu_1\ldots\mu_n}(k)\mathcal O_{\nu_1\ldots\nu_n}(-k)} ={\cal F}_{\mu_1\ldots\mu_n\nu_1\ldots\nu_n}(k,m_1,m_2)$ would be a tensor object. One can always decompose this tensor into a suitably constructed basis (using the vector $k_\mu$ and trivial unit tensors), and by projecting on the basic tensor structures, one can construct the spectral densities associated to each structure. The latter problem can then be reduced to that of the scalar case by contractions. For example, assume that the two-point function is a rank 2 tensor, $\mathcal{F}_{\mu\nu}(k)$. Each rank two tensor can be decomposed into its transversal and longitudinal part, according to
\begin{equation}\label{extra}
    \mathcal{F}_{\mu\nu}(k)= \mathcal{F}_T(k) \underbrace{\left(g_{\mu\nu}-\frac{k_\mu k_\nu}{k^2}\right)}_{P_{\mu\nu}(k)}+\mathcal{F}_L(k) \underbrace{\left(\frac{k_\mu k_\nu}{k^2}\right)}_{L_{\mu\nu}(k)}\,,
\end{equation}
so that knowledge of $\mathcal{F}_T(k)$, resp.~$\mathcal{F}_L(k)$, follows from the (scalar) quantity $P_{\mu\nu}(k)\mathcal{F}_{\mu\nu}(k)$, resp.~$L_{\mu\nu}(k)\mathcal{F}_{\mu\nu}(k)$. Using these scalar objects, one can the construct the spectral densities corresponding to the transversal and longitudinal piece of the two-point functionb. The here described procedure can be generalized to more complicated tensorial structures.

We thus wish to find the spectral representation \eqref{cutk-drel} of the integral \eqref{1-loop-mink}, and we shall therefore use the Cutkosky cut rules. In order to do so, we must assume that $m_1^2$ and $m_2^2$ are real, and that we are in Minkowski space time.

By cutting the lines comprising this $1$-loop diagram and, according to the rules, performing the replacement \cite{smatrix2,Peskin:1995ev}
\begin{align}
\label{cutk-cutrules-a}
 \frac{1}{(p^2-m^2_i)} \rightarrow 2\pi \theta(p^0)\delta(p^2-m^2_i)\,,
\end{align}
the spectral function is proportional to the discontinuity, $\rho(\tau) \propto \text{Disc } {\cal F}(\tau)=2\text{Im } {\cal F}(\tau)$, and it can be determined from
\begin{align}
\label{cutk-cutrules}
\text{Im } {\cal F} (k,m_1,m_2) = \frac 12 \int \frac{d^dp}{(2\pi)^d}\; \left[(2\pi)^2 \theta((k-p)^0)\delta((k-p)^2-m^2_1)\theta(p^0)\delta(p^2-m^2_2)f(p,k-p) \right]\,.
\end{align}
In order to evaluate (\ref{cutk-cutrules}) we will work in a frame where $k^{\mu} = (k^0, 0) = (E, 0)$. We have then
\begin{align}
\label{cutk-eval}
\text{Im } {\cal F} (E^2) = \frac 12 \int \frac{d^dp}{(2\pi)^{(d-2)}}\; \left[\theta(E-p^0)\delta((E-p^0)^2-\omega^2_{p,1})\theta(p^0)\delta((p^0)^2-\omega^2_{p,2}) f(p,k-p) \right]\,,
\end{align}
where $\omega_{p,i} = \sqrt{\vec{p}^2+ m^2_i}$. In the evaluation of this integral one must observe that the function $f(p,k-p)$ comes from possible derivatives in the operator ${\cal O}(k)$  and that it is a Lorentz scalar. It can thus only be a function of $E$, $m_1$ and $m_2$, which comes from the scalars $p^2 = m_2^2$, $(k-p)^2 = m_1^2$ and $p\cdot (k-p) = \frac12 (E^2 - m_1^2 - m_2^2)$. After the evaluation we can write the Lorentz invariant spectral function $\rho(\tau)$ by going to an arbitrary frame, whilst replacing $E^2 \rightarrow \tau$.

We are ultimately interested, however, in the case where the masses involved are complex and the momenta are in Euclidean space. We then consider expressions like (\ref{cutk-eval}) as functions of real masses and Minkowski momenta and will analytically continue\footnote{A certain care is needed when speaking about ``analytic continuation'', as we are considering a function depending on two variables, viz.~$m_1^2$ and $m_2^2$.} it to complex mass values in Euclidean momentum space. We want to draw attention here that, if we would immediately assume $m_1^2$ and $m_2^2$ to be complex, we cannot apply the cut rules. We cannot put an $i$-particle physically on-shell, or mathematically spoken, \eqref{cutk-cutrules-a} does not make much sense for $m_i^2\in\mathbb{C}$ while $p^2\in\mathbb{R}$.

\subsection{Naive application of the cut rules}
We will now illustrate the procedure just described by computing spectral functions in different dimensionalities for the two-point function \eqref{1-loop-mink} with $f(p,k-p)=1$. This corresponds to the toy model introduced in \cite{Baulieu:2009ha}. We are thus interested in finding a spectral representation of the form
\begin{equation}\label{setup0}
F_{d}(k^2)=\int_{\tau_0}^{\infty} d\tau\frac{\rho_d(\tau)}{\tau+k^2}
\end{equation}
for the following Euclidean two-point function
\begin{equation}\label{setup1}
    F_{d}(k^2)=\int \frac{d^dp}{(2\pi)^d}\frac{1}{p^2+m_1^2}\frac{1}{(p-k)^2+m_2^2}\,,
\end{equation}
with conjugate masses parametrized as
\begin{equation}\label{setup2}
    m_1^2=a+ib\,,\qquad m_2^2=a-ib\,.
\end{equation}
We assume that\footnote{This in order to avoid tachyon instabilities. For $b=0$, this is immediately clear. In the context of the model of \cite{Dudal:2008sp}, $a$ itself also corresponds to the mass of another degree of freedom, as such $a$ is supposed to be positive.} $a\geq 0$, and without loss of generality we may take $b\geq 0$.

We shall therefore first use the Minkowski formula \eqref{cutk-eval}, and integrate over $p^0$ in (\ref{cutk-eval}), obtaining
\begin{align}
\label{cutk-eval2}
\text{Im } {\cal F} (E^2) = \frac 12 \int \frac{d^{(d-1)}p}{(2\pi)^{(d-2)}}\; \frac{1}{2\omega_{p,2}}\left[\theta(E-\omega_{p,2})\delta((E-\omega_{p,2})^2-\omega^2_{p,1})\right]\,.
\end{align}
The integrand depends only on $\vec{p}^2$ so that we can write
\begin{align}
\label{cutk-eval3}
\text{Im } {\cal F} (E^2) = \frac 12 \frac{1}{(2\pi)^{(d-2)}} \frac{2\pi^{\frac{d-1}{2}}}{\Gamma(\frac{d-1}{2})}\int^{\infty}_0 d|\vec{p}|\; |\vec{p}|^{(d-2)}\;  \frac{1}{2\omega_{p,1}} \frac{1}{2\omega_{p,2}}\delta(E-\omega_{p,1}-\omega_{p,2})\,,
\end{align}
where we also took into account the theta function. The remaining delta function can be cast in a more convenient form using the property $\delta(g(\vec{p}^2)) = \frac{1}{|g'(\vec{p}_0^2)|}\delta(\vec{p}^2-\vec{p}^2_0) = \frac{1}{2|\vec{p}_0||g'(\vec{p}_0^2)|}\delta(|\vec{p}|-|\vec{p}_0|)$ where $\vec{p}_0^2$ is such that $g(\vec{p}_0^2) = 0$. In our case, $\vec{p}_0^2$ satisfies
\begin{align}
\label{cutk-delta}
g(\vec{p}_0^2) \equiv E-\sqrt{\vec{p}_0^2+ m^2_1}-\sqrt{\vec{p}_0^2+ m^2_2} = 0\,.
\end{align}
The integral \eqref{cutk-eval3} can now be readily evaluated
\begin{align}
\label{cutk-eval4}
\text{Im } {\cal F} (E^2) = \frac{1}{(2\pi)^{(d-2)}}\frac{\pi^{\frac{d-1}{2}}}{\Gamma(\frac{d-1}{2})}\frac{|\vec{p}_0|^{d-3}}{4E} \,,
\end{align}
where
\begin{align}
\label{cutk-delta1}
|\vec{p}_0| = \frac{\sqrt{(E^2-m_1^2-m_2^2)^2 - 4m_1^2 m_2^2}}{2E} \,,
\end{align}
Having done this, we can now pass to the Euclidean case with the masses given by \eqref{setup2}, by evaluating this expression \eqref{cutk-delta1},
\begin{align}
\label{cutk-delta2}
|\vec{p}_0| = \frac{\sqrt{E^4- 4b^2 - 4aE^2}}{2E}\,.
\end{align}
The expression \eqref{cutk-eval4} can then worked out further for the various dimensions. For $d=2$ we obtain
\begin{align}
\label{cutk-eval-2D}
\text{Im } {\cal F}_{d=2} (E^2) = \frac{1}{2\sqrt{E^4- 4b^2 - 4aE^2}}\,.
\end{align}
We eventually conclude that
\begin{align}
\label{cutk-eval-2D-b}
\rho_{d=2} (\tau) = \frac{1}{2\pi} \frac{1}{\sqrt{\tau^2- 4b^2 - 4a\tau}}\,,
\end{align}
using $E^2 \rightarrow \tau$ and the equivalence $\rho(\tau) = \frac 1\pi \text{Im} {\cal F} (\tau)$.

Analogously, for $d=3$ and $d=4$, we find
\begin{align}
\label{cutk-eval-3D}
\text{Im } {\cal F}_{d=3} (E^2) = \frac{1}{8E}  &\longrightarrow \rho_{d=3}(\tau) = \frac{1}{8\pi} \frac{1}{\sqrt{\tau}}\,,\\
\text{Im } {\cal F}_{d=4} (E^2) = \frac{1}{16\pi} \sqrt{1-\frac{4b^2}{E^4}-\frac{4a}{E^2}}  &\longrightarrow \rho_{d=4}(\tau) = \frac{1}{(4\pi)^2} \sqrt{1-\frac{4b^2}{\tau^2}-\frac{4a}{\tau}}\,.
\end{align}
The threshold $\tau_0$ is in all cases given by
\begin{equation}\label{thres}
\tau_0=(m_1+m_2)^2 = 2\left(a + \sqrt{a^2 + b^2}\right)\,.
\end{equation}

\sect{Survey of the Widder Stieltjes inversion operator}
In this section, we shall first refresh the concept of the Stieltjes integral transformation, and then discuss how the inverse transformation can be found.

Let us assume a function $F(x)$, defined for $x>0$, and implicitly define another function $\rho(t)$ by means of
\begin{equation}\label{2}
    F(x)=\int_0^{+\infty}dt\frac{\rho(t)}{t+x}\,,
\end{equation}
where we assume that the latter integral exists. This operation defines the Stieltjes integral transformation \cite{stieltjes1,stieltjes2}. Upon comparing \eqref{cutk-drel} and \eqref{2}, it is clear that a Stieltjes representation \eqref{2} of the function $F(x)$ is nothing else than the K\"all\'{e}n-Lehmann representation if $F(x)$ would be a two-point function in Euclidean space. If $\rho(t)=0$ for $t<A$, $A$ sets the threshold $\tau_0$.

The spectral density $\rho(t)$ can be reconstructed from \cite{stieltjes1,stieltjes2}
\begin{equation}\label{3}
\rho(t)=\lim_{n\to +\infty} (-1)^{n+1} \frac{1}{(n!)^2} \p_t^{n}\left[t^{2n+1}\p_t^{n+1}F(t)\right]\,.
\end{equation}
Let us first prove this statement along the lines of \cite{stieltjes1}, as the tools of the proof shall turn out to be useful for our later analysis too. We start with
\begin{eqnarray}\label{4}
    (-1)^{n+1} \frac{1}{(n!)^2} \p_t^{n}\left[t^{2n+1}\p_t^{n+1}F(t)\right]&=&(-1)^{n+1} \frac{1}{(n!)^2} \p_t^{n}\left[t^{2n+1}\p_t^{n+1}\int_0^{+\infty}du\frac{\rho(u)}{t+u}\right]\nonumber\\
    &=&\frac{n+1}{n!}\p_t^{n}\left[t^{2n+1}\int_0^{+\infty}du\frac{\rho(u)}{(t+u)^{n+2}}\right]\nonumber\\
    &=&\frac{n+1}{n!} \int_0^{+\infty} du  \rho(u)u^n \p_t^{n}\left[\frac{t^{2n+1}}{u^n(t+u)^{n+2}}\right]\,.
\end{eqnarray}
It seems to be more involved to compute the latter derivative, but we can use a nice trick to do so \cite{stieltjes1}. Defining
\begin{eqnarray}\label{5}
g(t,u)=\frac{t^{2n+1}}{u^n(t+u)^{n+2}}\,,
\end{eqnarray}
we notice that this is a homogenous function of order $-1$, as for any $\ell>0$,
\begin{eqnarray}\label{6}
g(\ell t,\ell u)=\ell^{-1}g(t,u)\,.
\end{eqnarray}
Deriving w.r.t. $\ell$ and setting $\ell=1$ leads to the Euler characterization of homogenous functions. Specifically,
\begin{eqnarray}\label{7}
t\p_t g+u\p_u g &=& -g\,,
\end{eqnarray}
or for $t>0$,
\begin{eqnarray}\label{8}
\p_t g&=&-\p_u\left(\frac{u}{t}g\right)\,.
\end{eqnarray}
If $g(t,u)$ is homogenous of order $-1$, so is $\frac{u}{t}g(t,u)$, hence we can iteratively employ \eqref{8} to compute
\begin{eqnarray}\label{9}
\p_t^{n}g(t,u)&=&(-1)^n \p_u^{n} \left[\frac{u^n}{t^n}g(t,u)\right]= (-1)^n t^{n+1} \p_u^{n}\frac{1}{(t+u)^{n+2}}=t^{n+1}\frac{(2n+1)!}{(n+1)!}\frac{1}{(t+u)^{2n+2}}\,.
\end{eqnarray}
We are thus lead to
\begin{eqnarray}\label{10}
\eqref{4}&=& \frac{(2n+1)!}{(n!)^2}\int_0^{+\infty} du \frac{\rho(u)}{u} \left(\frac{ut}{(u+t)^2}\right)^{n+1}=\frac{(2n+1)!}{(n!)^2}\int_0^{+\infty} du \frac{\rho(u)}{u} e^{(n+1)h(u,t)}\,,
\end{eqnarray}
with
\begin{equation}\label{h}
h(u,t)=\ln\frac{ut}{(u+t)^2}\,.
\end{equation}
In the limit $n\to\infty$,  the latter integral is ideally suited for a steepest descent evaluation, since
\begin{equation}
\left.\p_u h(u,t)\right|_{u=t}=0\,,\qquad \left.\p_u^2 h(u,t)\right|_{u=t}=-\frac{1}{2t^2}<0\,.
\end{equation}
Doing so, we find
\begin{eqnarray}\label{11}
    \lim_{n\to +\infty}\frac{(2n+1)!}{(n!)^2}\int_0^{+\infty} du \frac{\rho(u)}{u} e^{(n+1)h(u,t)}&=&\lim_{n\to +\infty}\frac{(2n+1)!}{(n!)^2} \frac{\rho(t)}{t} e^{(n+1)h(t,t)}\int_0^{+\infty}du e^{-(n+1)\frac{(u-t)^2}{4t^2}}\,,\nonumber\\
\end{eqnarray}
with
\begin{equation}
h(t,t)=-2\ln 2\,,
\end{equation}
and, for $n\to+\infty$,
\begin{eqnarray}\label{12}
    \int_0^{+\infty} du e^{-(n+1)\frac{(u-t)^2}{4t^2}}\to    \int_{-\infty}^{+\infty} du e^{-(n+1)\frac{(u-t)^2}{4t^2}}=2\frac{\sqrt{\pi}}{\sqrt{1+n}}t\,.
\end{eqnarray}
Recalling Stirling's formula
\begin{equation}\label{13}
    n!\to\sqrt{2\pi n}\left(\frac{n}{e}\right)^n\,,\qquad \textrm{for}\;n\to+\infty\,,
\end{equation}
we can write that
\begin{equation}\label{14b}
    \frac{(2n+1)!}{(n!)^2}\to\frac{1}{2\sqrt{\pi}}4^{n+1}\sqrt{n+1}\,,\qquad \textrm{for}\;n\to+\infty\,.
\end{equation}
We then find
\begin{eqnarray}\label{14}
\lim_{n\to +\infty}\frac{(2n+1)!}{(n!)^2} \frac{\rho(t)}{t} e^{(n+1)h(t,t)}\int_0^{\infty}du e^{-(n+1)\frac{(u-t)^2}{4t^2}}=\rho(t)\,,
\end{eqnarray}
hereby proving \eqref{3}.

Once having determined $\rho(t)$ via \eqref{3}, one can then define
\begin{equation}\label{15}
    F(z)=\int_0^{+\infty}dt\frac{\rho(t)}{t+z}
\end{equation}
for $z\in\mathbb{C}$, which is analytic, with the exception of a branch cut on the negative real axis, $z\leq 0$ \cite{stieltjes2}. Obviously, there will be no branch cut for $z\in[-\delta_2,-\delta_1]$ if $\rho(t)=0$ for $t\in[\delta_1,\delta_2]$, with $\delta_1>\delta_2\geq 0$.  Using Cauchy's formula, one can then see that it actually holds that
\begin{equation}\label{16}
\rho(t)=\frac{1}{2\pi i}\lim_{\epsilon\to 0^+}\left[F(-t-i\epsilon)-F(-t+i\epsilon)\right]\,,
\end{equation}
which corresponds to the discontinuity of $F(z)$ along the negative real axis. En route, this explains the dispersion relation \eqref{cutk-drel}.

Apparently, \eqref{16} would provide us a with a much easier way to compute the spectral density $\rho(t)$, but the difficulty is that in most cases, we do not know how to evaluate $F(z)$ from its original (integral) definition for $z\not\in \mathbb{R}_0^+$. In contrast with this, \eqref{3} only needs $F(z)$ for $z\in\mathbb{R}_0^+$. This is exactly what we need, since we can always evaluate the Euclidean momentum integrals defining a two-point function $F(k^2)$, in which case we can assume $k^2\in\mathbb{R}_0^+$. Afterwards, $F(z)$ is \emph{defined} by means of \eqref{15} for all $z\in\mathbb{C}$.

In principle, \eqref{15} also makes sense when $\rho(t)$ is a distribution. A basic example is $\rho(t)=\delta(t-m^2)$, which leads to the familiar Yukawa-propagator $F(k^2)=\frac{1}{k^2+m^2}$, which evidently has no branch cut.

\sect{A first application: two (positive) real masses}
As a warming up exercise, we shall treat here a well-known textbook example, and compute the spectral density in the case of 2 real masses. This was already treated in \cite{Itzykson:1980rh}, albeit in Minkowski space. We shall work in Euclidean space, and we start from \eqref{setup1}. We introduce a Feynman parameter $x$, yielding
\begin{equation}\label{int2}
F_d(k^2)=    \int_0^1dx\int \frac{d^dp}{(2\pi)^d}\frac{1}{[x((k-p)^2+m_1^2)+(1-x)(p^2+m_2^2)]^2}\,.
\end{equation}
The substitution $q=p-xk$ gives
\begin{equation}\label{int4}
F_d(k^2)=    \int_0^1dx\int \frac{d^dq}{(2\pi)^d}\frac{1}{(q^2+\Delta^2)^2}\,,
\end{equation}
with
\begin{equation}\label{int5}
\Delta^2=xk^2+xm_1^2+m_2^2-xm_2^2-x^2k^2\,.
\end{equation}
Usage of the standard Euclidean space formula
\begin{equation}\label{int6}
    \int\frac{d^dq}{(2\pi)^d}\frac{1}{(q^2+\Delta^2)^n}=\frac{1}{(4\pi)^{d/2}}(\Delta^2)^{d/2-n}\frac{\Gamma(n-d/2)}{\Gamma(n)}
\end{equation}
leads to
\begin{equation}\label{int7}
F_d(k^2)= \frac{1}{(4\pi)^{d/2}}(\Delta^2)^{d/2-2}\Gamma(2-d/2)\,.
\end{equation}
For the rest of this section, we shall mainly concentrate ourselves on the $d=2$ case. We are thus interested in
\begin{equation}\label{ex1}
    F_{d=2}(k^2)=\frac{1}{4\pi}\int_0^1 \frac{dx}{x(1-x)k^2+x(m_1^2-m_2^2)+m_2^2}\,.
\end{equation}
Setting $t=k^2$ and dropping the irrelevant prefactor, we may focus on
\begin{equation}\label{ex1bis}
    F(t)=\int_0^1 \frac{dx}{x(1-x)k^2+x(m_1^2-m_2^2)+m_2^2}\,.
\end{equation}
Subsequently,
\begin{equation}\label{ex2}
    \p_t^{n+1}F(t)= \p_t^{n+1} \int_0^1 \frac{dx}{x(1-x)}\frac{1}{t+\alpha}=(-1)^{n+1}(n+1)!\int_0^1\frac{dx}{x(1-x)}\frac{1}{(t+\alpha)^{n+2}}\,,
\end{equation}
where we temporarily set $\alpha\equiv\alpha(x)=\frac{x(m_1^2-m_2^2)+m_2^2}{x(1-x)}$. Doing so, we have
\begin{equation}\label{ex3}
    \rho(t)=\int_0^1\frac{dx}{x(1-x)}\lim_{n\to+\infty}\frac{n+1}{n!} \p_t^n\left(\frac{t^{2n+1}}{(t+\alpha)^{n+2}}\right)\,.
\end{equation}
We already computed the derivative appearing in the r.h.s. of \eqref{ex3}, reusing the result \eqref{9} yields
\begin{eqnarray}\label{ex4}
    \rho(t)&=&\lim_{n\to+\infty}\frac{(2n+1)!}{(n!)^2} \int_0^1\frac{dx}{x(1-x)}\frac{\alpha^n t^{n+1}}{(t+\alpha)^{2n+2}}=\lim_{n\to+\infty}\frac{(2n+1)!}{(n!)^2} \int_0^1\frac{dx}{x(1-x)} \frac{1}{\alpha} e^{(n+1)h(\alpha,t)}\,.\nonumber\\
\end{eqnarray}
We notice the great power of this formulation, as the spectral density can now be obtained from a steepest descent evaluation of the Feynman parameter integral \eqref{ex4}. It is useful to perform the substitution $x=\frac{1}{1+y}$ to rewrite \eqref{ex4} as
\begin{eqnarray}\label{ex5}
    \rho(t)&=&\lim_{n\to+\infty}\frac{(2n+1)!}{(n!)^2} \int_0^{+\infty}\frac{dy}{(1+y)(m_1^2+ym_2^2)} e^{(n+1)h(\alpha,t)}\,,
\end{eqnarray}
with
\begin{equation}\label{ex6}
    \alpha\equiv\alpha(y)=\frac{1+y}{y}(m_1^2+ym_2^2)\,.
\end{equation}
We are now ready to search for the maxima of $h(y,t)=h(\alpha(y),t)$ for $y\in[0,\infty]$. Solving $\p_y h(y,t)=0$ gives
\begin{eqnarray}\label{ex7}
    y_1~&=&~\frac{m_1}{m_2}\,,\qquad y_2~=~\frac{t-m_1^2-m_2^2+\sqrt{-4m_1^2m_2^2+(-t+m_1^2+m_2^2)^2}}{2m_2^2}\,,\nonumber\\
        y_3~&=&~-\frac{m_1}{m_2}\,,\qquad y_4~=~\frac{t-m_1^2-m_2^2-\sqrt{-4m_1^2m_2^2+(-t+m_1^2+m_2^2)^2}}{2m_2^2}\,.
\end{eqnarray}
We can discriminate between a few cases.
\begin{enumerate}
\item $0<t< (m_1-m_2)^2$\\
In this case, the root appearing in \eqref{ex7} exists. However, only $y_1>0$. We have
\begin{equation}\label{ex8}
    \left.\p_y^2 h\right|_{y=y_1}=\frac{2m_2^3(t-(m_1+m_2)^2)}{m_1(m_1+m_2)^2(t+(m_1+m_2)^2)}\equiv -\beta^2<0\,.
\end{equation}
We can then look at the behaviour of the integral for $n\to+\infty$,
\begin{eqnarray}\label{ex9}
 \int_0^{+\infty}\frac{dy}{(1+y)(m_1^2+ym_2^2)} e^{(n+1)h(\alpha,t)}\propto \frac{1}{\sqrt{n+1}}e^{(n+1)h(y_1,t)}\,.
\end{eqnarray}
Using
\begin{equation}\label{ex10}
    h(y_1,t)=\ln\frac{t(m_1+m_2)^2}{(t+(m_1+m_2)^2)^2}\,,
\end{equation}
we find
\begin{eqnarray}\label{ex10}
 \int_0^{+\infty}\frac{dy}{(1+y)(m_1^2+ym_2^2)} e^{(n+1)h(\alpha,t)}\propto \frac{1}{\sqrt{n+1}}\left(\frac{t(m_1+m_2)^2}{(t+(m_1+m_2)^2)^2}\right)^{n+1}\,.
\end{eqnarray}
Using \eqref{14b}, we conclude that $\rho(t)=0$ as $0\leq\frac{t(m_1+m_2)^2}{(t+(m_1+m_2)^2)^2}< \frac{1}{4}$.

\item $(m_1-m_2)^2<t<(m_1+m_2)^2$\\
In this interval, the roots appearing in \eqref{ex7} are not real, so again only $y_1$ is relevant. The rest of the reasoning remains valid, so we conclude again that $\rho(t)=0$.

\item $t>(m_1+m_2)^2$\\
In this case, the roots are again real-valued. Looking at \eqref{ex8}, we notice that $y_1$ now corresponds to a minimum, so we can disregard it. Only $y_2$ and $y_4$ are relevant, as both are positive and correspond to local maxima. We shall not write down some intermediate results, as they are quite lengthy. We set
\begin{equation}\label{ex8}
    \left.\p_y^2 h\right|_{y=y_2}=-\beta_2^2<0\,,\qquad    =\left.\p_y^2 h\right|_{y=y_4}=-\beta_4^2<0\,.
\end{equation}
It then suffices to notice that
\begin{equation}\label{ex10}
    h(y_2,t)=h(y_3,t)=-2\ln 2\,,
\end{equation}
which is exactly the value we need to find a nonvanishing finite contribution to the integral. In particular,
\begin{eqnarray}\label{ex11}
 \rho(t)=\frac{1}{(1+y_2)(m_1^2+y_2m_2^2)}\frac{\sqrt{2}}{2\beta_2}+\frac{1}{(1+y_4)(m_1^2+y_4m_2^2)}\frac{\sqrt{2}}{2\beta_4}\,,
\end{eqnarray}
which simplifies to
\begin{eqnarray}\label{ex12}
 \rho(t)=\frac{2}{\sqrt{-4m_1^2m_2^2+(t-m_1^2-m_2^2)^2}}
\end{eqnarray}
after a bit of algebra.
\end{enumerate}
At the end of the day, we find that
\begin{equation}\label{c1a}
F_{d=2}(k^2)=\int_{(m_1+m_2)^2}^{+\infty}dt\frac{\rho(t)}{t+k^2}\,,\qquad\text{with }\rho(t)=\frac{1}{2\pi\sqrt{-4m_1^2m_2^2+(t-m_1^2-m_2^2)^2}}\,.
\end{equation}
It is easily verified that $\rho(t)\geq 0$ for $t\geq (m_1+m_2)^2$. These results are in full agreement with those obtained from the Cutkosky rules, see also \cite{Itzykson:1980rh}, given that we transform from Minkowski to Euclidean space.

\sect{A comment about divergent spectral integrals}
Setting $d=4$ in \eqref{int7}, we clearly stumble upon a divergence $\propto \Gamma(0)$. In order to deal with finite quantities, one usually turns to a suitably subtracted spectral representation. In particular, if we formally set
\begin{equation}\label{exb1a}
    G(x)=\int_{0}^{\infty} dt\frac{\rho(t)}{t+x}~(=\infty)
\end{equation}
for a divergent\footnote{We assume here that the divergency is due to the large $t$-behaviour of $\rho(t)$, something which is usually the case.} quantity $G(x)$, it is clear that a suitable number of derivatives w.r.t.~$x$ will render us with a finite result,
\begin{equation}\label{exb1b}
 H(x)=   \frac{\p^N}{(\p x)^N}G(x)=(-1)^N\int_{0}^{\infty} dt\frac{\rho(t)}{(t+x)^{N+1}}<\infty\,,
\end{equation}
leading to a finite subtracted spectral representation after $N$ integrations from $0$ to $x$,
\begin{equation}\label{exb1c}
    G^\text{sub}(x)=G(x)-\ldots -\frac{x^{n-1}}{(N-1)!}\frac{\p^{N-1} G}{\p x^{N-1}}(0)=(-1)^N x^N\int_{0}^{\infty} dt\frac{\rho(t)}{t^N(t+x)}<\infty\,.
\end{equation}
In order to know the spectral representation corresponding to $G(x)$, it looks preferential to study the finite function $H(x)$, \eqref{exb1b}, and bring it into a spectral representation of the form
\begin{equation}\label{exb1d}
 H(x)=   (-1)^N\int_{0}^{\infty} dt\frac{\rho(t)}{(t+x)^{N+1}}\,,
\end{equation}
from which we can also read off the desired spectral density $\rho(t)$ of the original function $G(x)$. Using the techniques of Section 2, it is not difficult to show that\footnote{We may always take $n\geq N$.}
\begin{equation}\label{exb1e}
\rho(t)=\lim_{n\to +\infty} (-1)^{n+1-N} \frac{N!}{(n!)^2} \p_t^{n}\left[t^{2n+1}\p_t^{n+1-N}H(t)\right]\,,
\end{equation}
a formula which allows to compute the spectral density $\rho(t)$ from knowledge of the quantity $H(x)$, and this without encountering any infinities at any time.

In practice, glancing again at \eqref{int7}, we would have
\begin{equation}\label{exb2}
    \frac{\p F_{d=4}(k^2)}{\p k^2}=-\frac{1}{16\pi^2}\int_0^1 dx\frac{1}{k^2+\frac{x(m_1^2-m_2^2)+m_2^2}{x(1-x)}}\,,
\end{equation}
and upon comparing with \eqref{ex1}, we notice that these are almost identical. Only the ``prefactor function'' of the $\frac{1}{t+\alpha}$ will be different. In particular, this means that the saddle point equation etc for $d=4$ remains the same as for $d=2$, just as almost all the rest of the derivation. Only during the last step in computing the spectral density, we shall notice a difference, as only there the ``prefactor function'' plays a role.

The case $d=3$ would need to be worked out from the beginning, but similar tricks as for $d=2$ or $d=4$ can be applied. This can be easily appreciated as a systematics in the derivatives of $\frac{1}{\sqrt{t+\alpha}}$ exists, the latter being the basic form that will appear in the $d=3$ case.

\sect{A second application: two complex conjugate masses}
We now come to the most interesting application of this paper. We reconsider the integral \eqref{setup1}, rewritten in the form \eqref{ex1}, but now we immediately assume that the occurring mass scales are complex conjugate, and given by \eqref{setup2}. It can be checked that the ``trick-inspired'' approaches of \cite{Baulieu:2009ha} to construct the spectral density, do not longer work out in the case that $b\neq0$. We wish to confirm that the results \eqref{cutk-eval-2D-b} and \eqref{thres} obtained via a blind trust in the Cutkosky approach, are actually correct.

We start our analysis from the first line of \eqref{ex4}.  We have
\begin{eqnarray}\label{c1}
    \rho(t)&=&\lim_{n\to+\infty}\frac{(2n+1)!}{(n!)^2} \int_0^1\frac{dx}{x(1-x)}\frac{\alpha^n t^{n+1}}{(t+\alpha)^{2n+2}}\,,
\end{eqnarray}
now with
\begin{equation}\label{c2}
\alpha\equiv\alpha(x)=\frac{x(m_1^2-m_2^2)+m_2^2}{x(1-x)}=\frac{a}{x(1-x)}+ib\frac{2x-1}{x(1-x)}\,.
\end{equation}
In principle, one might try to do a steepest descent evaluation of the integral \eqref{c1}, keeping in mind that this time one would need to deform the integration contour to follow a line of steepest descent through the equivalent of the saddle points \eqref{ex7} in the complex $x$-plane. This looks as a very complicated exercise, instead we shall follow a somewhat different reasoning. Let us first notice that \eqref{c1} defines a real function, as
\begin{eqnarray*}\label{c3}
    \int_0^1\frac{dx}{x(1-x)}\frac{\alpha^n t^{n+1}}{(t+\alpha)^{2n+2}}&=&   \int_0^{1/2}\frac{dx}{x(1-x)}\frac{1}{\alpha}\left(\frac{\alpha t}{(t+\alpha)^{2}}\right)^{n+1}+\int_{1/2}^1\frac{dx}{x(1-x)}\frac{1}{\alpha}\left(\frac{\alpha t}{(t+\alpha)^{2}}\right)^{n+1}\nonumber\\
    &=&\int_0^{1/2}\frac{dx}{x(1-x)}\left[\frac{1}{\alpha}\left(\frac{\alpha t}{(t+\alpha)^{2}}\right)^{n+1}\right]+\int_0^{1/2}\frac{dx}{x(1-x)}\left[\frac{1}{\alpha^*}\left(\frac{\alpha^* t}{(t+\alpha^*)^{2}}\right)^{n+1}\right]\,,
\end{eqnarray*}
via the substitution $x\to1-x$ in the 2nd integral. We can thus say
\begin{eqnarray}\label{c4}
    \rho(t)&=&2\mathrm{Re}[R(t)]\,,
\end{eqnarray}
and it suffices to study
\begin{eqnarray}\label{c5}
    R(t)=\lim_{n\to+\infty}\frac{(2n+1)!}{(n!)^2} \int_0^{1/2}\frac{dx}{x(1-x)}\frac{1}{\alpha}\left(\frac{\alpha t}{(t+\alpha)^{2}}\right)^{n+1}\,.
\end{eqnarray}
It would be beneficial if we could use $\alpha$ as (complex) integration variable, whereby we integrate along the contour\footnote{We do not necessarily refer to a \emph{closed} contour here!} $\gamma$, which has the parameter representation
\begin{equation}\label{c6}
   \gamma: x\in[0,1/2]\to \alpha(x)=\frac{a}{x(1-x)}+ib\frac{2x-1}{x(1-x)}\,.
\end{equation}
This $\gamma$ will start from a point at complex infinity in the right lower half plane and will end on the real axis at $\alpha=4a$. Inverting gives
\begin{equation}\label{c7}
    x=\frac{-2ib+\alpha\pm\sqrt{-4b^2-4a\alpha+\alpha^2}}{2\alpha}\,.
\end{equation}
A careful examination learns that the minus sign is the appropriate choice, given that $\alpha(x)$ lives in the lower half $\alpha$-plane for $x\in[0,1/2]$. We can subsequently reexpress \eqref{c5} as
\begin{eqnarray}\label{c8}
    R(t)=\lim_{n\to+\infty}\frac{(2n+1)!}{(n!)^2} \int_\gamma\frac{-d\alpha}{\sqrt{-4b^2+\alpha(-4a+\alpha)}}\frac{1}{\alpha}\left(\frac{\alpha t}{(t+\alpha)^2}\right)^{n+1}
\end{eqnarray}
after a little algebra. For the further analysis, it is easier to switch the orientation of the contour $\gamma$, so than we can write
\begin{eqnarray}\label{c9}
    R(t)
    &=&\lim_{n\to+\infty}\frac{(2n+1)!}{(n!)^2} \int_\Gamma\frac{d\alpha}{\sqrt{-4b^2+\alpha(-4a+\alpha)}}\frac{1}{\alpha}e^{(n+1)\ln h(\alpha,t)}\,,
\end{eqnarray}
with $\Gamma$ as shown in Figure 1, and $h(\alpha,t)$ given by \eqref{h}.
\begin{figure}[H]
\begin{center}
\includegraphics[width=8cm]{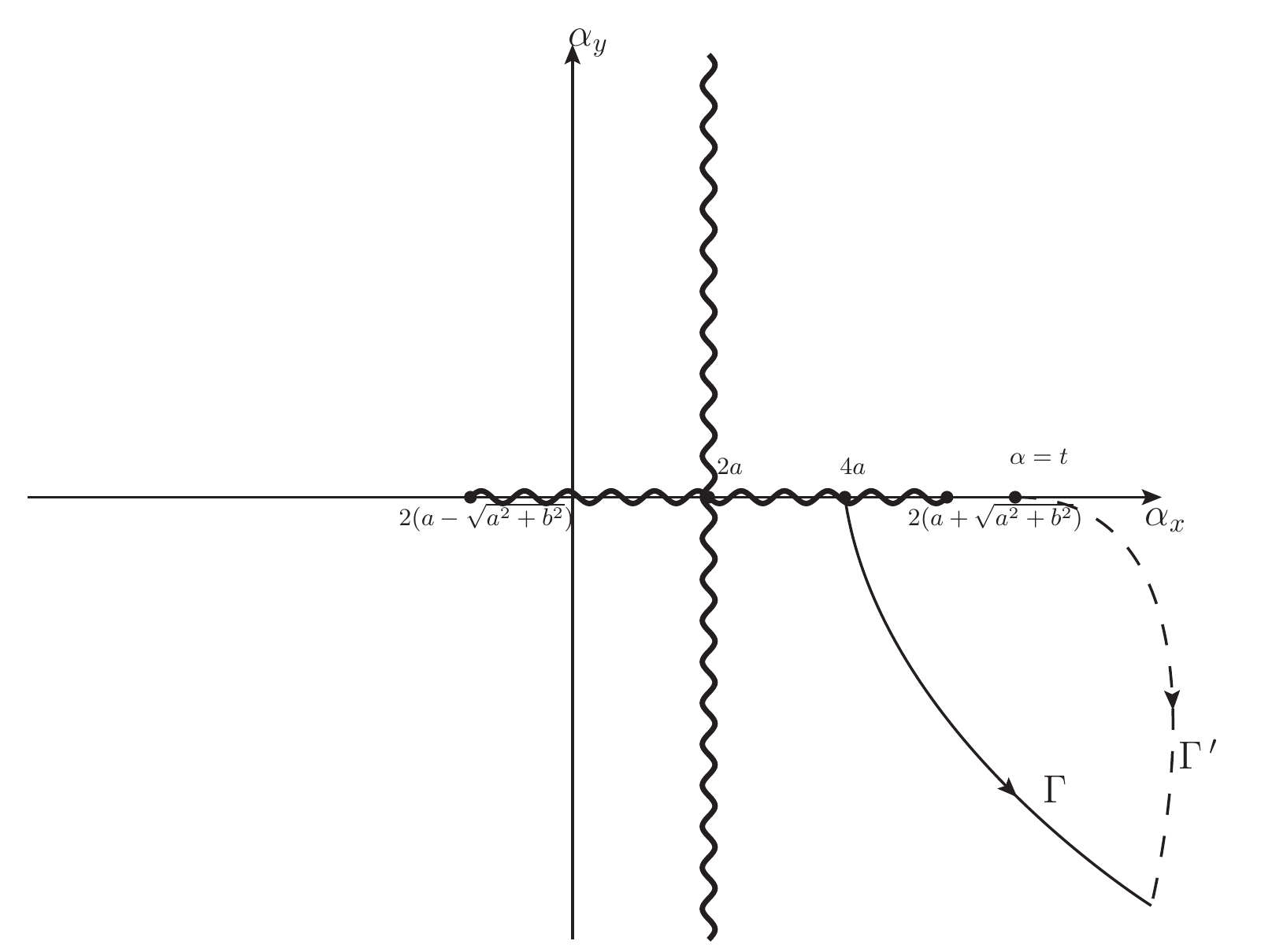} \label{fig1}
     \end{center}
  \caption{}
\end{figure}

The presence of the square root in \eqref{c9} complicates the situation a bit. Setting $\alpha=\alpha_x+i\alpha_y$ learns there is a ``crucifix-cut'' given by
\begin{equation}\label{c10}
    \left\{\begin{array}{l}
      \alpha_x  =2a\,, \alpha_y \in\mathbb{R} \\
      \alpha_x\in[2(a-\sqrt{a^2+b^2}),2(a+\sqrt{a^2+b^2})]\,,\alpha_y=0\,.
    \end{array}\right.
\end{equation}
Our goal is now to compute \eqref{c9} by means of a steepest descent approach. We can use previously gained knowledge, from which we learn that $h(\alpha,t)$ has a saddle point at $\alpha=t$. In the vicinity of the saddle point, we may write
\begin{equation}\label{c11}
    \ln h(\alpha,t)=-2\ln 2 -\frac{(\alpha-t)^2}{4t^2}+\ldots\,.
\end{equation}
We can now consider a few cases.
\begin{enumerate}
\item $t>2(a+\sqrt{a^2+b^2})$\\ Since everything is analytic in the considered region of the complex plane\footnote{The presence of the $\ln$ is of no concern for the analyticity, since $n\in\mathbb{N}$.}, we can deform our contour $\Gamma$ into $\Gamma'$, which shares its begin and end point with $\Gamma$, such that $\Gamma'$ passes through the saddle point. We have to choose the orientation of the contour $\Gamma'$ in the vicinity of the saddle point in such a way that the imaginary part of $\ln h$ remains constant. From \eqref{c11}, it is easily seen that this is the case when we take $\alpha$ real for a while until we have passed $t$, and then we can let $\Gamma'$ bend over to let it flow to the correct end point at infinity\footnote{As the function vanishes at infinity, we can also move the end point at infinity, if desired.}. Doing so, we find
    \begin{eqnarray}\label{c12}
    &&\lim_{n\to\infty}\int_\Gamma\frac{d\alpha}{\sqrt{-4b^2+\alpha(-4a+\alpha)}}\frac{1}{\alpha}e^{(n+1)\ln h(\alpha,t)}\nonumber\\&=&\lim_{n\to\infty}e^{-2(n+1)\ln2}\frac{1}{\sqrt{-4b^2+t(-4a+t)}}\frac{1}{t}\int_{-\infty}^{+\infty} d\alpha_x e^{-(n+1)}\frac{(\alpha_x-t)^2}{4t}
    \nonumber\\&=&\lim_{n\to\infty}\frac{1}{4^{n+1}}\frac{2\sqrt{\pi}}{\sqrt{1+n}}\frac{1}{\sqrt{-4b^2+t(-4a+t)}}
    \end{eqnarray}
    where we made use of \eqref{12}. Combining this result with \eqref{14b} gives
\begin{eqnarray}\label{c13}
    R(t)&=&\frac{1}{\sqrt{-4b^2+t(-4a+t)}}\,.
\end{eqnarray}
    Hence,
\begin{eqnarray}\label{c14}
    \rho(t)&=&\frac{2}{\sqrt{-4b^2+t(-4a+t)}}\,.
\end{eqnarray}
\item $4a<t<2(a+\sqrt{a^2+b^2})$\\
In this region, we can still deform the contour $\Gamma$ into an appropriate $\Gamma'$ in a completely similar way, as we do not have to cross any cut in order to do so. Since we are below the cut, we find after a completely analogous reasoning as in the first case
\begin{eqnarray}\label{c15}
    R(t)&=&\frac{i}{\sqrt{4b^2+t(4a-t)}}\,,
\end{eqnarray}
so that now
\begin{eqnarray}\label{c16}
    \rho(t)&=&0\,.
\end{eqnarray}

\item $2a<t<4a$\\
At first sight, we can repeat the analysis of subcase 2. This reasoning is however flawed. A basic ingredient of the steepest descent approach is that $\mathrm{Re}(\ln h(\alpha,t))$  reaches its maximum along the (deformed) contour at the value of the saddle point, in this case $\alpha=t$. If we have to first stretch our contour $\Gamma$ into the left direction to pick up the saddle point, we are violating this assumption, as it can be checked. The grey region in Figure 2 displays the region in the complex $\alpha$-plane we cannot cross with our contour.
 \begin{figure}[H]
\begin{center}
\includegraphics[width=8cm]{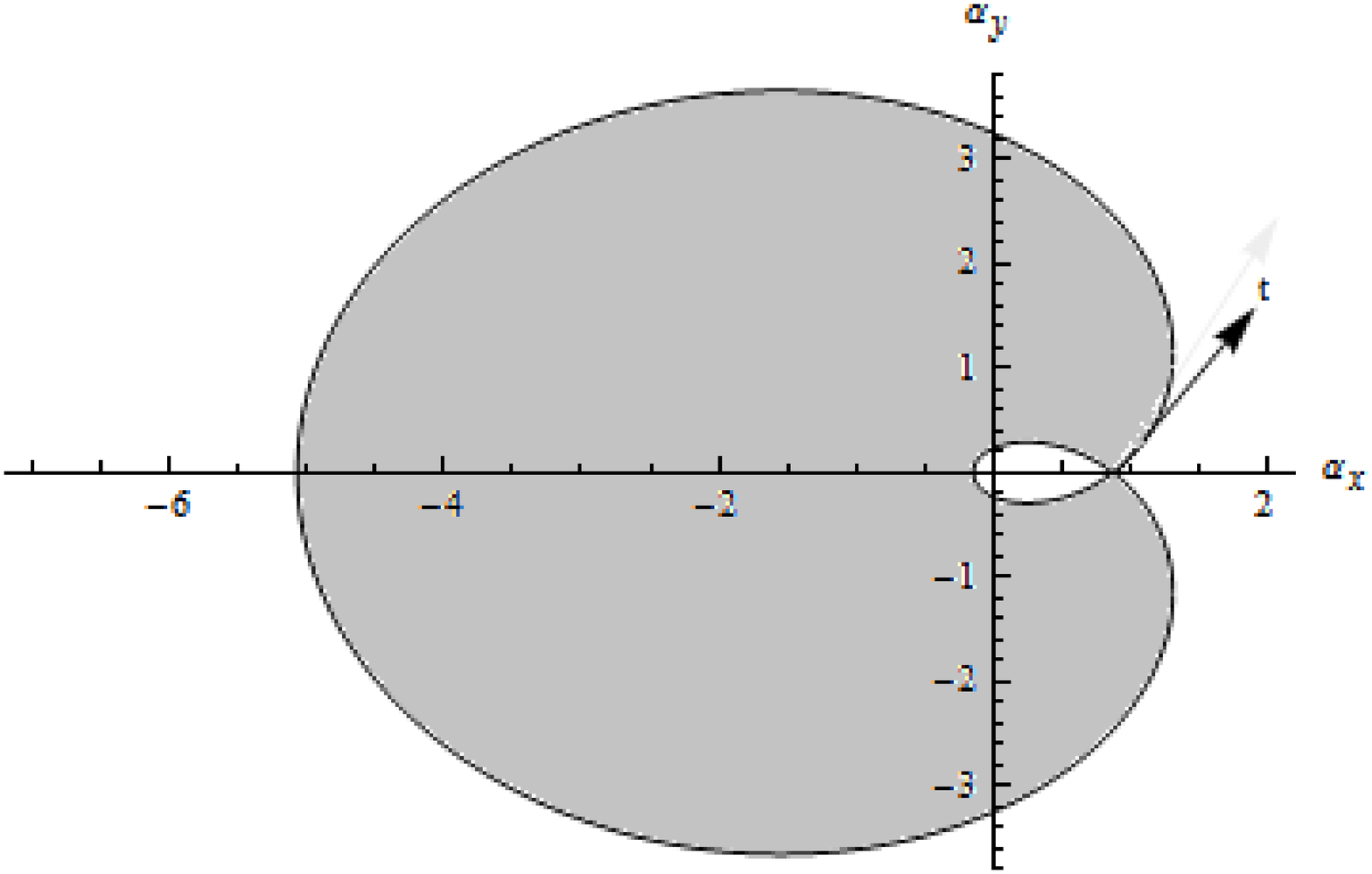} \label{fig2}
     \end{center}
  \caption{}
\end{figure}
The boundaries of these curves are defined by
\begin{equation}\label{bound}
   \alpha_y=\pm \sqrt{7 t^2-4 \sqrt{t^3 (3 t-2 \alpha_x)}-2 t \alpha_x-\alpha_x^2}\,,\qquad \alpha_y=\pm\sqrt{7 t^2-2 t \alpha_x-\alpha_x^2+4 \sqrt{3 t^4-2 t^3 \alpha_x}}\,.
\end{equation}
If $t>4a$, the begin point of $\Gamma$ lies within the white blob, and we can (only) follow the $\alpha_x$-axis, pick up the saddle point at $\alpha_x=t$, leave this region and go to infinity. If however we start outside this region, there is no way that we can go through the saddle point and go back to infinity without self-crossing the contour. Said otherwise, the steepest descent approach is inapplicable in this case.

\item $0\leq t <2a$\\
In this region, we seem to be in even more serious trouble, as we can even no longer deform our contour in a useful way, due to the vertical branch cut at $\alpha=2a$.

\item $0\leq t <4a$\\
Let us now present a combined study of the two previous cases, based on a different approach. Using the triangle inequality for integrals, we have
\begin{eqnarray}\label{c17b}
|R(t)|&=&\lim_{n\to+\infty}\frac{(2n+1)!}{(n!)^2}\left|\int_0^{1/2}\frac{dx}{x(1-x)}\frac{\alpha^n t^{n+1}}{(t+\alpha)^{2n+2}}\right|\nonumber\\&\leq& \lim_{n\to+\infty}\frac{(2n+1)!}{(n!)^2}\int_0^{1/2}\frac{dx}{x(1-x)}\left|\frac{\alpha^n t^{n+1}}{(t+\alpha)^{2n+2}}\right|\nonumber\\
&=&\lim_{n\to+\infty}\frac{(2n+1)!}{(n!)^2}\int_0^{1/2} \frac{dx}{x(1-x)}\frac{1}{\left|\alpha\right|}\left(\frac{t\sqrt{a^2+b^2(1-2x)^2}x(1-x)}{(a-tx(1-x))^2+b^2(1-2x)^2}\right)^{n+1}\,.
\end{eqnarray}
We consequently observe that the (positive) function
\begin{equation}\label{c18}
    j(x,t)=\frac{t\sqrt{a^2+b^2(1-2x)^2}x(1-x)}{(a-tx(1-x))^2+b^2(1-2x)^2}
\end{equation}
has maxima at the solutions of
\begin{equation}\label{c19}
    \frac{\p j}{\p x}=    \frac{\p j}{\p t}=0\,,
\end{equation}
namely,
\begin{equation}\label{c20}
(x,t)=(0,0)\,,\qquad (x,t)=(1,0)\,,\qquad (x,t)=(1/2,4a)\,,
\end{equation}
with
\begin{equation}\label{c21}
j(0,0)=0\,,\qquad j(1,0)=0\,,\qquad j(1/2,4a)=\frac{1}{4}\,.
\end{equation}
On the interval $(x,t)\in [0,1/2]\times[0,2a]$, this implies that, for a certain $\zeta$,
\begin{equation}\label{c22}
0\leq j(x,t)\leq\zeta<\frac{1}{4}\,,
\end{equation}
and using this we can majorate \eqref{c17b} further as
\begin{eqnarray}\label{c23}
|R(t)|&\leq&\lim_{n\to+\infty}\frac{(2n+1)!}{(n!)^2}\zeta^{n+1}\underbrace{\int_0^{1/2} \frac{dx}{x(1-x)}\frac{1}{\left|\alpha\right|}}_{\textrm{number independent of }n}\,.
\end{eqnarray}
Once more using \eqref{14b}, we then simply find
\begin{eqnarray}\label{c24}
|R(t)|&\leq&0\,,
\end{eqnarray}
as $\zeta<\frac{1}{4}$. A fortiori, $R(t)=0$, and thus also $\rho(t)=0$.

\end{enumerate}
Combining all information gathered so far, we have shown that
\begin{equation}\label{c25}
    F_{d=2}(k^2)=\int_{2(a+\sqrt{a^2+b^2})}^{+\infty}dt\frac{\rho(t)}{t+k^2}\,,\qquad \textrm{with }\rho(t)=\frac{1}{2\pi\sqrt{-4b^2+t(-4a+t)}}\,,
\end{equation}
whereby we can observe that $\rho(t)\geq 0 $ for $t\geq2(a+\sqrt{a^2+b^2})$.

As a check on the final result \eqref{c25}, we notice that its $b\to0$ limit coincides with the $m_1^2\to m_2^2\to a$ limit of the earlier obtained result \eqref{c1a}.

For the general case, we could of course explicitly compute the spectral integral of \eqref{c25}, as well as the original Feynman parameter integral \eqref{ex1}, and verify if both results are the same. We find it however more instructive for the reader to display in Figure 3 both results for the explicit example $m_1^2=1+2i$, $m_2^2=1-2i$, making their equivalence clearly visible.\\
\begin{figure}[H]
  \begin{center}
  \subfigure[]{\includegraphics[width=7cm]{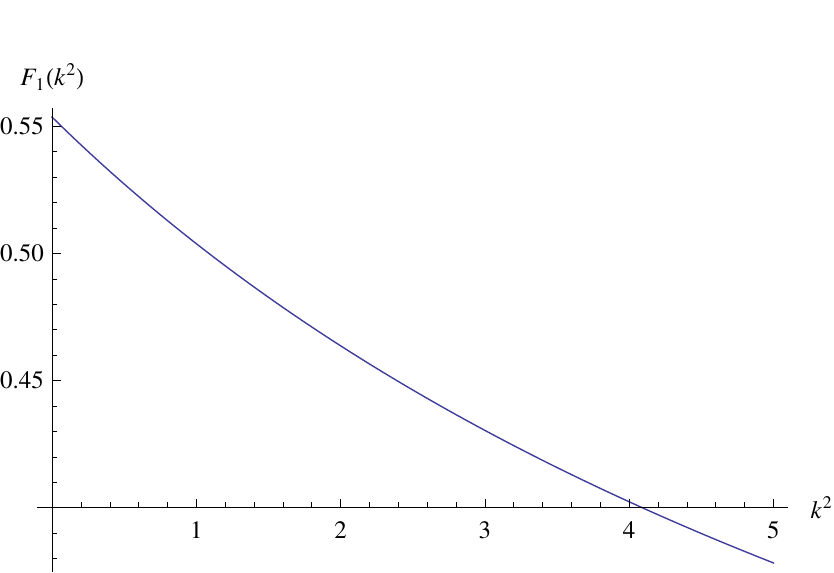} \label{fig2a}}
    \hspace{1cm}
 \subfigure[]{\includegraphics[width=7cm]{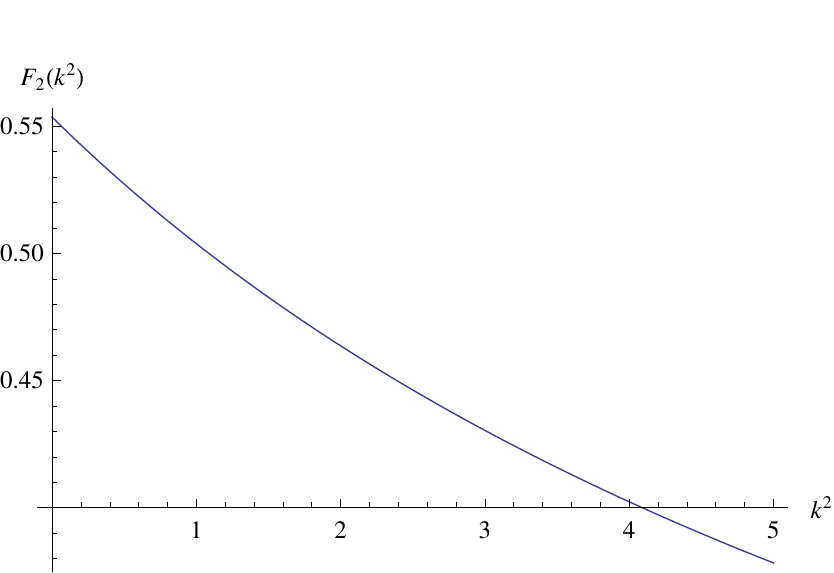}\label{fig2b}}
     \end{center}
  \caption{Plots of $\displaystyle F_1(k^2)=\int_0^1 \frac{dx}{x(1-x)t+2ibx+a-ib}$ and $\displaystyle F_2(k^2)=\int_{2(a+\sqrt{a^2+b^2})}^{+\infty}dt\frac{\rho(t)}{t+k^2}$ for $a=1$, $b=2$.}
\end{figure}
Similar conclusions can be reached for $d=3$ or $d=4$. In all cases, we thus find perfect agreement between the results obtained by application of the Cutkosky cut rules, and those obtained by using the Widder Stieltjes formalism.

\sect{Concluding remarks}
We have demonstrated the usefulness of the Stieltjes integral transform to compute the K\"all\'{e}n-Lehmann spectral density of (Euclidean) two-point functions, even in cases where the validity of the usually employed Cutkosky rules is not clear. The main recipes are given in Sections 4 and 5, in particular in expression \eqref{3} which gives the spectral density $\rho(t)$ in terms of the two-point function $F(k^2)$, when the latter is known for Euclidean momenta $k^2\geq 0$.

We have paid particular attention to the case of Gribov-like propagators, which entails the presence of propagators with 2 complex conjugate masses. It turns out that, at least at one loop, this case can we worked out analytically to the end using a steepest descent approach. Our main conclusion is that the results, obtained by the cut rules for real masses and in Minkowski space, may be continued to Euclidean space, while also the two real masses can formally be replaced by two complex conjugate values. This kind of results already found use in recent works like \cite{Capri:2010pg,Dudal:2010cd}, and have now been proven to be correct.

However, the inversion formula \eqref{3} is also valid in cases where the cut rules are unavailable. We recall that many works have been devoted to the study of the quantum equations of motion (Schwinger-Dyson formalism), giving also estimates for the gluon propagator, amongst other quantities \cite{Alkofer:2000wg,Roberts:1994dr,Fischer:2008uz,Fischer:2008uzb,Fischer:2008uzc,Boucaud:2010gr}. In the deep infrared and ultraviolet, an analytical approach is possible. In the intermediate momentum regime, however, numerical solutions are in order due to the complexity of the system. The numerical outcome can then fitted by well-guessed functional forms. Let us mention two major examples
\begin{eqnarray}\label{fit1}
    D(k^2)&=&\frac{1}{k^2+m^2(k^2)}\,,\quad m^2(k^2)~=~ \frac{m_0^4}{k^2+m_0^2}\left[\frac{\ln\frac{k^2+f(k^2,m_0^2)}{\Lambda^2}}{\ln\frac{f(0,m_0^2)}{\Lambda^2}}\right]^{-3/5}\,,\nonumber\\
    f(k^2,m_0^2)&=&\rho_1 m_0^2+\rho_2\frac{m_0^4}{q^2+m_0^2}\,,\quad \rho_1=-\frac{1}{2}\,,\quad\rho_2=\frac{5}{2}\,,\quad m_0=612\,\text{MeV}\,, \quad\Lambda=645\,\text{MeV}\,,
\end{eqnarray}
which represents the so-called ``massive'' \cite{Fischer:2008uz} (also known as ``decoupling'' \cite{Fischer:2008uzc}) solution, or
\begin{eqnarray}\label{fit2}
D(k^2)&=& Z_{I,II}(k^2)\left[\frac{\alpha(0)}{1+k^2/\Lambda^2}+\frac{4\pi}{\beta_0}\frac{k^2}{k^2+\Lambda^2}\left(\frac{1}{\ln k^2/\Lambda^2}-\frac{\Lambda^2}{k^2-\Lambda^2}\right)\right]^{\frac{13}{22}}\,,\nonumber\\
\beta_0&=&\frac{11N}{3}\,,\quad\alpha(0)=\frac{8.915}{N}\,,\quad \Lambda=710\,\text{MeV}\,,
\end{eqnarray}
with
\begin{eqnarray}\label{fit2b}
Z_I(k^2)= w_I \frac{(k^2)^{2\kappa}}{(k^2)^{2\kappa}+(\Lambda^2)^{2\kappa}}\,, \quad \text{or}\, Z_{II}(k^2)= w_{II} \left(\frac{k^2}{k^2+\Lambda^2}\right)^{2\kappa}\,,\quad\kappa\approx 0.595\,,
\end{eqnarray}
for the so-called ``scaling'' solution \cite{Alkofer:2003jj}.

It is evidently out of the question to apply the standard cut rules with the foregoing propagators, since they are not even of the form $\frac{1}{p^2+m^2}$, so the replacement rule \eqref{cutk-cutrules-a} loses its meaning. The theory of the (Stieltjes) inversion could however be applied, albeit perhaps rather in a numerical fashion, employing stable approximations for the infinite numbers of derivatives appearing in formula \eqref{3}. Let us also mention here that in many cases, the concrete application of the Cutkosky rules become intractable beyond one loop, due to the highly complicated phase space integrals.

In the particular case of bound states, like a glueball or meson, which can crucially depend on a viable input for the gluon propagator, one might imagine to use propagators of the type \eqref{fit1} or \eqref{fit2} to compute, in one approximation or another, suitable two-point functions, to obtain consequently an estimate for the spectral density, and then eventually use the latter as input for whatever method one likes to employ to find estimates for e.g.~bound state masses. One might think about employing Laplacian sum rules as in \cite{Narison:2002pw} or techniques based on Pad\'{e} approximants as in \cite{Dudal:2010cd,Bhagwat:2007rj}, to name only a few approaches to the bound state problem whereby the spectral density enters.

We end by mentioning that \eqref{fit1} or \eqref{fit2} are quite different in the deep infrared as \eqref{fit1} tends to a strictly positive constant for $k^2\to0$, while \eqref{fit2} vanishes in the same limit. The former scenario is recovered on the lattice \cite{Cucchieri:2007md,Cucchieri:2007mdb,Cucchieri:2007mdc}. One could nevertheless wonder if both solutions could reproduce more or less the same physics, as preliminary investigated in \cite{Blank:2010pa} in a specific example and approximation. We hope to come back to this issue in the future, using the tools developed in this paper. The content of Section 5 should be of particular interest, as also the propagators \eqref{fit1} or \eqref{fit2} will lead to divergent spectral integrals.

\section*{Acknowledgments}
We wish to thank S.~P.~Sorella, N.~Temme and N.~Vandersickel for useful discussions. D.~Dudal is supported by the Research-Foundation
Flanders (FWO Vlaanderen), while M.~S.~Guimar\~{a}es acknowledges financial support by FAPERJ, Funda{\c{c}}{\~{a}}o de Amparo {\`{a}} Pesquisa do Estado do Rio de Janeiro. D.~Dudal acknowledges the hospitality at the UERJ where this work was initiated and finished.


\begin{thebibliography}{99}
\bibitem{smatrix1}
R.~J.~Eden, P.~V.~Landshoff, D.~I.~Olive and J.~C.~Polkinghorne, \emph{The analytic S-matrix}, Cambridge at the University Press (1966).

\bibitem{smatrix2}
I.~T.~Todorov, \emph{Analytic properties of Feynman diagrams in quantum field theory}, Pergamon Press (1971).

\bibitem{Itzykson:1980rh}
C.~Itzykson and J.~B.~Zuber, \emph{Quantum Field Theory}, New York, USA: McGraw-Hill (1980) (International Series In Pure and Applied Physics).

\bibitem{Peskin:1995ev}
M.~E.~Peskin and D.~V.~Schroeder, \emph{An Introduction To Quantum Field Theory}, Reading, USA: Addison-Wesley (1995).

\bibitem{Cutkosky:1960sp}
R.~E.~Cutkosky, J.\ Math.\ Phys.\  {\bf 1} (1960) 429.

\bibitem{Narison:2002pw}
S.~Narison, Camb.\ Monogr.\ Part.\ Phys.\ Nucl.\ Phys.\ Cosmol.\  {\bf 17 } (2002)  1.

\bibitem{Shifman:1978bx}
M.~A.~Shifman, A.~I.~Vainshtein and V.~I.~Zakharov, Nucl.\ Phys.\  B {\bf 147} (1979) 385.

\bibitem{Shifman:1978by}
M.~A.~Shifman, A.~I.~Vainshtein and V.~I.~Zakharov, Nucl.\ Phys.\  B {\bf 147} (1979) 448.

\bibitem{Alkofer:2000wg}
R.~Alkofer and L.~von Smekal, Phys.\ Rept.\  {\bf 353} (2001) 281.

\bibitem{Roberts:1994dr}
C.~D.~Roberts and A.~G.~Williams, Prog.\ Part.\ Nucl.\ Phys.\  {\bf 33} (1994) 477.

\bibitem{Roberts:2007ji}
C.~D.~Roberts, Prog.\ Part.\ Nucl.\ Phys.\  {\bf 61} (2008) 50.

\bibitem{Cucchieri:2007md}
A.~Cucchieri and T.~Mendes, PoS {\bf LAT2007 } (2007)  297.

\bibitem{Cucchieri:2007mdb}
A.~Cucchieri and T.~Mendes, Phys.\ Rev.\ Lett.\  {\bf 100 } (2008)  241601.

\bibitem{Cucchieri:2007mdc}
I.~L.~Bogolubsky, E.~M.~Ilgenfritz, M.~Muller-Preussker {\it et al.}, PoS {\bf LAT2007 } (2007)  290.

\bibitem{Fischer:2008uz}
A.~C.~Aguilar, D.~Binosi and J.~Papavassiliou, Phys.\ Rev.\  D {\bf 78} (2008) 025010.

\bibitem{Fischer:2008uzb}
D.~Binosi and J.~Papavassiliou, Phys.\ Rept.\  {\bf 479} (2009) 1.

\bibitem{Fischer:2008uzc}
C.~S.~Fischer, A.~Maas and J.~M.~Pawlowski, Annals Phys.\  {\bf 324} (2009) 2408.

\bibitem{Dudal:2008sp}
D.~Dudal, J.~A.~Gracey, S.~P.~Sorella, N.~Vandersickel and H.~Verschelde, Phys.\ Rev.\  D {\bf 78} (2008) 065047.

\bibitem{Dudal:2010tf}
D.~Dudal, O.~Oliveira and N.~Vandersickel, Phys.\ Rev.\  D {\bf 81} (2010) 074505.

\bibitem{Gracey:2010cg}
J.~A.~Gracey, Phys.\ Rev.\  D {\bf 82} (2010) 085032.

\bibitem{Boucaud:2010gr}
Ph.~Boucaud, M.~E.~Gomez, J.~P.~Leroy, A.~Le Yaouanc, J.~Micheli, O.~Pene and J.~Rodriguez-Quintero, Phys.\ Rev.\  D {\bf 82} (2010) 054007.

\bibitem{Tissier:2010ts}
M.~Tissier and N.~Wschebor, Phys.\ Rev.\  D {\bf 82} (2010) 101701.

\bibitem{Furui:2004cx}
S.~Furui and H.~Nakajima, Phys.\ Rev.\  D {\bf 70} (2004) 094504.

\bibitem{Maas:2009se}
A.~Maas, Phys.\ Lett.\  B {\bf 689} (2010) 107.

\bibitem{Kondo:2009ug}
K.~I.~Kondo, Phys.\ Lett.\  B {\bf 678} (2009) 322.

\bibitem{Iritani:2009mp}
T.~Iritani, H.~Suganuma and H.~Iida, Phys.\ Rev.\  D {\bf 80} (2009) 114505.

\bibitem{Bogolubsky:2009dc}
I.~L.~Bogolubsky, E.~M.~Ilgenfritz, M.~Muller-Preussker and A.~Sternbeck, Phys.\ Lett.\  B {\bf 676} (2009) 69.

\bibitem{Oliveira:2010xc}
O.~Oliveira and P.~Bicudo, arXiv:1002.4151 [hep-lat].

\bibitem{Osterwalder:1974tc}
K.~Osterwalder and R.~Schrader, Commun.\ Math.\ Phys.\   {\bf 31} (1973) 83.

\bibitem{Osterwalder:1974tcb}
K.~Osterwalder and R.~Schrader, Commun.\ Math.\ Phys.\  {\bf 42} (1975) 281.

\bibitem{Silva:2006bs}
P.~J.~Silva and O.~Oliveira, PoS {\bf LAT2006} (2006) 075.

\bibitem{Bowman:2007du}
P.~O.~Bowman,  U.~M.~Heller, D.~B.~Leinweber, M.~B.~Parappilly, A.~Sternbeck, L.~von Smekal, A.~G.~Williams and J.-b.~Zhang, Phys.\ Rev.\  D {\bf 76} (2007) 094505.

\bibitem{Gribov:1977wm}
V.~N.~Gribov, Nucl.\ Phys.\  B {\bf 139} (1978) 1.

\bibitem{Zwanziger:1989mf}
D.~Zwanziger, Nucl.\ Phys.\  B {\bf 323} (1989) 513.

\bibitem{Zwanziger:1989mfb}
D.~Zwanziger, Nucl.\ Phys.\  B {\bf 399}  (1993) 477.

\bibitem{Baulieu:2009ha}
L.~Baulieu, D.~Dudal, M.~S.~Guimaraes, M.~Q.~Huber, S.~P.~Sorella, N.~Vandersickel and D.~Zwanziger, Phys.\ Rev.\  D {\bf 82} (2010) 025021.

\bibitem{Stingl:1985hx}
M.~Stingl, Phys.\ Rev.\  D {\bf 34} (1986) 3863  [Erratum-ibid.\  D {\bf 36} (1987) 651].

\bibitem{Stingl:1994nk}
M.~Stingl, Z.\ Phys.\  A {\bf 353} (1996) 423.

\bibitem{Capri:2010pg}
M.~A.~L.~Capri, A.~J.~Gomez, M.~S.~Guimaraes, V.~E.~R.~Lemes, S.~P.~Sorella and D.~G.~Tedesco, arXiv:1009.3062 [hep-th].

\bibitem{Dudal:2010cd}
D.~Dudal, M.~S.~Guimaraes and S.~P.~Sorella, arXiv:1010.3638 [hep-th].

\bibitem{stieltjes1}
D.~V.~Widder, \emph{An introduction to transform theory}, Academic Press (1971).

\bibitem{stieltjes2}
L.~Debnath and D.~Bhatta, \emph{Integral transforms and their applications}, Taylor and Francis Group (2007).

\bibitem{Alkofer:2003jj}
R.~Alkofer, W.~Detmold, C.~S.~Fischer and P.~Maris, Phys.\ Rev.\  D {\bf 70} (2004) 014014.

\bibitem{Aguilar:2009ke}
A.~C.~Aguilar and J.~Papavassiliou, Phys.\ Rev.\  D {\bf 81} (2010) 034003.

\bibitem{Bhagwat:2007rj}
M.~S.~Bhagwat, A.~Hoell, A.~Krassnigg, C.~D.~Roberts and S.~V.~Wright, Few Body Syst.\  {\bf 40} (2007) 209.

\bibitem{Blank:2010pa}
M.~Blank, A.~Krassnigg and A.~Maas, arXiv:1007.3901 [hep-ph].


\end{thebibliography}
\end{document}